# The Structural Modelling of Operational Risk via Bayesian inference: Combining Loss Data with Expert Opinions

**Pavel V. Shevchenko**
CSIRO Mathematical and Information Sciences, Sydney, Locked Bag 17, North Ryde, NSW, 1670, Australia.    e-mail: Pavel.Shevchenko@csiro.au

**Mario V. Wüthrich**
ETH Zürich, Department of Mathematics, HG G 32.5, CH-8092, Zurich, Switzerland.
e-mail: wueth@math.ethz.ch

**Abstract**
To meet the Basel II regulatory requirements for the Advanced Measurement Approaches, the bank's internal model must include the use of internal data, relevant external data, scenario analysis and factors reflecting the business environment and internal control systems. Quantification of operational risk cannot be based only on historical data but should involve scenario analysis. Historical internal operational risk loss data have limited ability to predict future behaviour moreover, banks do not have enough internal data to estimate low frequency high impact events adequately. Historical external data are difficult to use due to different volumes and other factors. In addition, internal and external data have a survival bias, since typically one does not have data of all collapsed companies. The idea of scenario analysis is to estimate frequency and severity of risk events via expert opinions taking into account bank environment factors with reference to events that have occurred (or may have occurred) in other banks. Scenario analysis is forward looking and can reflect changes in the banking environment. It is important to not only quantify the operational risk capital but also provide incentives to business units to improve their risk management policies, which can be accomplished through scenario analysis. By itself, scenario analysis is very subjective but combined with loss data it is a powerful tool to estimate operational risk losses. Bayesian inference is a statistical technique well suited for combining expert opinions and historical data. In this paper, we present examples of the Bayesian inference methods for operational risk quantification.

# 1  Introduction

Under the Basel II requirements, see BIS (2005), a bank intending to use the Advanced Measurement Approaches (AMA) for the quantification of operational risk should demonstrate accuracy of internal models within Basel II risk cells (eight business lines times seven risk types) relevant to the bank. To meet the regulatory requirements, the model should make use of internal data, relevant external data, scenario analysis and factors reflecting the business environment and internal control systems. Basel II defines operational risk as the risk of loss resulting from inadequate or failed internal processes, people and systems or from external events. This definition includes legal risk, but excludes strategic and reputational risk. There are various aspects of operational risk modelling, see for e.g. Chavez-Demoulin, Embrechts and Nešlehová (2006) or Cruz (2004). Under the Loss Distribution Approach (LDA) for AMA, banks should quantify distributions for frequency and severity of operational losses for each risk cell (business line/event type) over a one year time horizon. The commonly used model for the annual loss in a single risk cell (business line/event type) is a compound process

$$Z = \sum_{i=1}^{N} X_i , \qquad (1)$$

where $N$ is the annual number of events modelled as a random variable from some discrete distribution (typically Poisson) and $X_i$, $i = 1,...,N$, are severities of the events modelled as independent random variables from a continuous distribution. Frequency $N$ and severities $X_i$ are assumed independent. Note that the independence assumed here is conditional on distribution parameters. Estimation of the annual loss distribution by modelling frequency and severity of losses is a well-known actuarial technique, see e.g. Klugman, Panjer, and Willmot (1998). It is also used to model solvency requirements in the insurance industry, see e.g. Sandström (2006), Wüthrich (2006).

   The estimation of frequency and severity distributions is a challenging task, especially for low frequency high impact losses. The banks internal data (usually truncated below approximately US$20,000) are available typically over several years and contain few (or none) high impact low frequency losses. The external data (losses experienced by other banks) are available through third party databases, but these are difficult to use directly due to different volumes and other factors. Typically, external data are available above US$1million. Moreover, the data have a survival bias as typically the data of all collapsed companies are not available. It is difficult to estimate distributions using these data only. It is also clear that this estimation is backward looking and has limited ability to predict the future due to a constantly changing banking environment. For example, assume that a new policy was introduced in the bank, aiming to decrease the operational risk losses. Then it cannot be captured in the model based on the loss data only. As another example, assume that the risk has a true arrival rate of 1/100. A bank started to collect data two years ago and by chance this risk event occurred within this period. Formally, applying loss data approach, the arrival rate of this risk might be estimated as 1/2, which is clearly overestimated, however, it is important to take this event into account.

It is very important to have scenario analysis incorporated into the model. In fact, it is mandatory to include scenario analysis into the model to meet the regulatory requirements. Scenario analysis is a process undertaken by banks to identify risks; analyse past events experienced internally and by other banks (including near miss losses); consider current and planned controls in the banks; etc. Usually, it involves workshops and templates to identify weaknesses, strengths and other factors. As a result some rough quantitative assessment of risk frequency and severity distributions is obtained from expert opinions. By itself, scenario analysis is very subjective and should be combined (supported) by the actual loss data analysis.

In practice, ad-hoc procedures are often used by banks to combine internal data, external data and expert opinions. For example:

- Fit distribution to the combined samples of internal and external data.
- Estimate event arrival rates $\lambda_{ext}$ and $\lambda_{int}$ implied by external and internal data, and combine them as $w\lambda_{int} + (1-w)\lambda_{ext}$ using expert specified (or calculated by ad hoc procedure) weight $w$ to estimate frequency distribution.
- Estimate distribution as a weighted sum $w_1 F_{SA}(X) + w_2 F_I(X) + (1 - w_1 - w_2) F_E(X)$, where $F_{SA}(X)$, $F_I(X)$, and $F_E(X)$ are the distributions identified by scenario analysis, internal data, and external data respectively, using ad-hoc calculated weights $w_1$ and $w_2$.

Bayesian inference is a statistical technique well suited to incorporate expert opinions into data analysis. There is a broad literature covering Bayesian inference and its applications for the insurance industry as well as other areas. For a good introduction to the Bayesian inference method, see Berger (1985), or for the closely related methods of credibility theory, see Bühlmann and Gisler (2005). The method allows for structural modelling where expert opinions are incorporated into the analysis via specifying distributions (so-called prior distributions) for model parameters. These are updated by the data as they become available. At any point in time, the expert may reassess the prior distributions, given the availability of new information (for example when new policy control is introduced), that will incorporate this information into a model. In our experience, this technique is rarely used for operational risk, although briefly mentioned in Cruz (2002). Below we describe this technique within the context of operational risk and provide several examples of its application for operational risk quantification.

## 2  Bayesian inference

Consider a random vector of observations $\mathbf{X} = (X_1, X_2, ..., X_n)$ whose density, for a given vector of parameters $\boldsymbol{\theta} = (\theta_1, \theta_2, ..., \theta_K)$, is $h(\mathbf{X}|\boldsymbol{\theta})$. In the Bayesian approach, both observations and parameters are considered to be random. Then Bayes' theorem can be formulated as

$$h(\mathbf{X}, \boldsymbol{\theta}) = h(\mathbf{X}|\boldsymbol{\theta})\pi(\boldsymbol{\theta}) = \hat{\pi}(\boldsymbol{\theta}|\mathbf{X})h(\mathbf{X}), \qquad (2)$$

where $\pi(\boldsymbol{\theta})$ is the density of parameters, a so-called prior distribution (typically, $\pi(\boldsymbol{\theta})$ depends on a set of further parameters that are called hyperparameters, omitted here for simplicity of notation); $\hat{\pi}(\boldsymbol{\theta}|\mathbf{X})$ is the density of parameters given observed data $\mathbf{X}$, a so-called posterior distribution; $h(\mathbf{X},\boldsymbol{\theta})$ is the joint density of observed data and parameters; $h(\mathbf{X}|\boldsymbol{\theta})$ is the density of observations for given parameters, and $h(\mathbf{X})$ is a marginal density of $\mathbf{X}$. The later can also be written as

$$h(\mathbf{X}) = \int h(\mathbf{X}|\boldsymbol{\theta})\pi(\boldsymbol{\theta})d\boldsymbol{\theta}. \qquad (3)$$

For simplicity of notation, we consider continuous $\pi(\boldsymbol{\theta})$ only. If $\pi(\boldsymbol{\theta})$ is a discrete distribution, then the integration in the above expression should be replaced with a summation $h(\mathbf{X}) = \sum h(\mathbf{X}|\boldsymbol{\theta})\pi(\boldsymbol{\theta})$.

The objective (in the context of operational risk) is to estimate the predictive distribution (frequency and severity) of a future observation $X_{n+1}$ conditional on all available information $\mathbf{X} = (X_1, X_2,..., X_n)$. Assume that conditionally, given parameters $\boldsymbol{\theta}$, $X_{n+1}$ and $\mathbf{X}$ are independent and $X_{n+1}$ has a density $f(X_{n+1}|\boldsymbol{\theta})$. Then the conditional density of $X_{n+1}$, given $\mathbf{X}$, is

$$f(X_{n+1}|\mathbf{X}) = \int f(X_{n+1}|\boldsymbol{\theta}) \times \hat{\pi}(\boldsymbol{\theta}|\mathbf{X})d\boldsymbol{\theta}. \qquad (4)$$

It is common to assume (and is assumed in all examples below) that $X_1, X_2,..., X_n, X_{n+1}$ are conditionally independent (given $\boldsymbol{\theta}$) and identically distributed. Using (2), the posterior distribution can be written as

$$\hat{\pi}(\boldsymbol{\theta}|\mathbf{X}) = h(\mathbf{X}|\boldsymbol{\theta})\pi(\boldsymbol{\theta})/h(\mathbf{X}). \qquad (5)$$

The distribution $h(\mathbf{X}|\boldsymbol{\theta})$ is a likelihood function of observations. Here, $h(\mathbf{X})$ plays the role of a normalization constant, thus, the posterior distribution can be viewed as a product of a prior knowledge with a likelihood function of observed data. In the context of operational risk we propose the following three steps:

- The prior distribution $\pi(\boldsymbol{\theta})$ should be estimated by scenario analysis (expert opinions with reference to external data).
- Then the prior distribution should be weighted with the observed data using formula (5) to get a posterior distribution $\hat{\pi}(\boldsymbol{\theta}|\mathbf{X})$.
- Formula (4) is then used to calculate the predictive distribution of $X_{n+1}$ given the observations $\mathbf{X}$.

The approach of the Bayesian estimates leads to optimal estimates in a sense that the mean square error of prediction is minimized, for more on this topic see e.g. Bühlmann and Gisler (2005).

**The iterative update procedure for priors**: If the observations $X_1, X_2, ..., X_n$ are conditionally (given $\boldsymbol{\theta}$) independent and identically distributed with density $f(.|\boldsymbol{\theta})$, then the likelihood function can be written as $h(\mathbf{X}|\boldsymbol{\theta}) = \prod_{i=1}^{n} f(X_i|\boldsymbol{\theta})$. Denote the posterior distribution calculated after $k$ observations as $\hat{\pi}_k(\boldsymbol{\theta}|X_1,...,X_k)$, then using (5), observe that

$$\hat{\pi}_k(\boldsymbol{\theta}|X_1,...,X_k) \propto \pi(\boldsymbol{\theta}) \prod_{i=1}^{k} f(X_i|\boldsymbol{\theta}) \propto \hat{\pi}_{k-1}(\boldsymbol{\theta}|X_1,...,X_{k-1}) \times f(X_k|\boldsymbol{\theta}). \qquad (6)$$

Hereafter, $\propto$ is used for statements with the relevant terms only. It is easy to see from (6), that the updating procedure which calculates the posteriors from priors can be done iteratively. Only the posterior distribution calculated after $k$-1 observations and the $k$-th observation are needed to calculate the posterior distribution after $k$ observations. Thus the loss history over many years is not required, making the model easier to understand and manage, and allowing experts to adjust the priors at every step. Formally, the posterior distribution calculated after $k$-1 observations can be treated as a prior distribution for the $k$-th observation. In practice, initially, we start with the prior distribution $\pi(\boldsymbol{\theta})$ identified by expert opinions and external data only. Then, the posterior distribution $\hat{\pi}(\boldsymbol{\theta}|\mathbf{X})$ is calculated, using (5), when actual data are observed. If there is a reason (for example, the new control policy introduced in a bank), then this posterior distribution can be adjusted by an expert and treated as the prior distribution for subsequent observations. Examples will be presented in the following sections.

## 3   Conjugate prior distributions

So called conjugate distributions are very useful in practice when Bayesian inference is applied. The precise definition, see e.g. Bühlmann and Gisler (2005), is:

**Definition**: Let $F$ denote the class of density functions $f(\mathbf{X}|\boldsymbol{\theta})$, indexed by $\boldsymbol{\theta}$. A class $U$ of prior densities $\pi(\boldsymbol{\theta})$ is said to be a conjugate family for $F$ if posterior density $\hat{\pi}(\boldsymbol{\theta}|\mathbf{X}) = f(\mathbf{X}|\boldsymbol{\theta})\pi(\boldsymbol{\theta})/f(\mathbf{X})$, where $f(\mathbf{X}) = \int f(\mathbf{X}|\boldsymbol{\theta})\pi(\boldsymbol{\theta})d\boldsymbol{\theta}$, is in the class $U$ for all $f \in F$ and $\pi \in U$.

Formally if the family $U$ contains all distribution functions then it is conjugate to any family $F$. However, to make a model useful in practice it is important that $U$ should be as small as possible while containing realistic distributions. Below we present $F$-$U$ conjugate pairs: Poisson-Gamma, LogNormal-Normal, Pareto-Gamma that are the most useful examples for modeling frequencies and severities in operational risk. Several other pairs (Binomial-Beta, Gamma-Gamma, Exponential-Gamma) can be found in e.g. Bühlmann and Gisler (2005). In all these cases, prior and posterior distributions have the same type and the posterior distribution parameters are easily calculated using the prior distribution parameters and observations (or recursively using (6)).

## 3.1 Poisson-Gamma (frequency modelling)

The Poisson distribution is often used for modelling the frequencies of operational risk losses. Suppose that conditionally, given $\lambda$, observations $\mathbf{N} = (N_1,...,N_n)$ are independent random variables from Poisson distribution, $Poisson(\lambda)$, with a density

$$f(N|\lambda) = e^{-\lambda}\frac{\lambda^N}{N!}, \quad \lambda \geq 0, \tag{7}$$

and prior distribution for $\lambda$ is Gamma distribution, $Gamma(\alpha, \beta)$ with a density

$$\pi(\lambda|\alpha,\beta) = \frac{(\lambda/\beta)^{\alpha-1}}{\Gamma(\alpha)\beta}\exp(-\lambda/\beta), \quad \lambda > 0, \alpha > 0, \beta > 0. \tag{8}$$

That is, $\lambda$ plays the role of $\boldsymbol{\theta}$ and $\mathbf{N}$ the role of $\mathbf{X}$ in (5). Given $\lambda$, $N_1,...,N_n$ are conditionally independent and their likelihood is given by

$$h(\mathbf{N}|\lambda) = \prod_{i=1}^{n} e^{-\lambda}\frac{\lambda^{N_i}}{N_i!}. \tag{9}$$

Then, using formula (5), the posterior distribution is

$$\hat{\pi}(\lambda|\mathbf{N}) \propto \frac{(\lambda/\beta)^{\alpha-1}}{\Gamma(\alpha)\beta}\exp(-\lambda/\beta)\prod_{i=1}^{n} e^{-\lambda}\frac{\lambda^{N_i}}{N_i!} \propto \lambda^{\hat{\alpha}-1}\exp(-\lambda/\hat{\beta}), \tag{10}$$

which is Gamma distribution, $Gamma(\hat{\alpha}, \hat{\beta})$, i.e. the same as the prior distribution with updated parameters $\hat{\alpha}$ and $\hat{\beta}$ given by:

$$\begin{aligned}\alpha \to \hat{\alpha} &= \alpha + \sum_{i=1}^{n} N_i, \\ \beta \to \hat{\beta} &= \beta/(1+\beta \times n).\end{aligned} \tag{11}$$

The expected number of events, given past observations, $E[N_{n+1}|\mathbf{N}]$, (which is a mean of the posterior distribution in this case) allows for a good interpretation, as follows:

$$E[N_{n+1} | \mathbf{N}] = E[\lambda | \mathbf{N}] = \hat{\alpha} \times \hat{\beta} = \beta \times \frac{\alpha + \sum_{i=1}^{n} N_i}{1 + \beta \times n} = w\overline{N} + (1-w)\lambda_0, \qquad (12)$$

where

$\overline{N} = \frac{1}{n}\sum_{i=1}^{n} N_i$ is the estimate of $\lambda$ using the observed counts only,

$\lambda_0 = \alpha \times \beta$ is the estimate of $\lambda$ using a prior distribution only (e.g. specified by expert),

$w = \frac{n}{n + 1/\beta}$ is the credibility weight in [0,1) used to combine $\lambda_0$ and $\overline{N}$.

As the number of years $n$ increases, the credibility weight $w$ increases and vice versa. That is, the more observations we have, the greater credibility weight we assign to the estimator based on the observed counts, while the lesser credibility weight is attached to the expert opinion estimate. Also, the larger the volatility of the expert opinion (larger $\beta$), the greater credibility weight is assigned to observations.

Recursive calculation of the posterior distribution is very simple. That is, consider observed annual counts $N_1, N_2,..., N_k,...$, where $N_k$ is the number of events observed in the $k$-th year. Assume that the prior distribution $\pi(\lambda | \alpha, \beta)$, $Gamma(\alpha, \beta)$, is specified initially, then the posterior distribution $\hat{\pi}_k(\lambda | N_1,..., N_k)$ after the $k$-th year is the Gamma distribution, $Gamma(\hat{\alpha}_k, \hat{\beta}_k)$, with $\hat{\alpha}_k = \alpha + \sum_{i=1}^{k} N_i$ and $\hat{\beta}_k = \beta/(1 + \beta \times k)$. Observe that,

$$\hat{\alpha}_k = \hat{\alpha}_{k-1} + N_k, \qquad \hat{\beta}_k = \hat{\beta}_{k-1}/(1 + \hat{\beta}_{k-1}). \qquad (13)$$

This leads to a very efficient recursive scheme, where the calculation of posterior distribution parameters is based on the most recent observation and parameters of posterior distribution calculated just before this observation.

### *3.2 LogNormal-Normal (severity modelling)*

The LogNormal distribution, $LN(\mu, \sigma)$ is often used to model the severity of operational risk losses (a numerical example is provided in Section 4). Suppose that conditionally, given $\mu$ and $\sigma$, observations $\mathbf{X} = (X_1,..., X_n)$ are independent random variables from $LN(\mu, \sigma)$ with a density

$$f(x | \mu, \sigma) = \frac{1}{x\sqrt{2\pi\sigma^2}} \exp\left(-\frac{(\ln x - \mu)^2}{2\sigma^2}\right). \qquad (14)$$

That is, $Y_i = \ln X_i$, $i = 1,...,n$, are distributed from the Normal distribution $N(\mu, \sigma)$. Assume that parameter $\sigma$ is known and the prior distribution for $\mu$ is the Normal distribution, $N(\mu_0, \sigma_0)$, with a density:

$$\pi(\mu \mid \mu_0, \sigma_0) = \frac{1}{\sigma_0 \sqrt{2\pi}} \exp\left(-\frac{(\mu - \mu_0)^2}{2\sigma_0^2}\right). \tag{15}$$

That is, $\mu$ plays the role of $\boldsymbol{\theta}$ in (5). The case of a conjugate joint prior for both $\mu$ and $\sigma$ unknown is considered in Appendix A. The likelihood of the observations (conditional on the parameters $\mu$ and $\sigma$) is

$$h(\mathbf{Y} \mid \mu, \sigma) = \prod_{i=1}^{n} \frac{1}{\sigma \sqrt{2\pi}} \exp\left(-\frac{(Y_i - \mu)^2}{2\sigma^2}\right). \tag{16}$$

Then, using formula (5), the posterior distribution can be written as

$$\hat{\pi}(\mu \mid \mathbf{Y}) \propto \frac{\exp\left(-\frac{(\mu - \mu_0)^2}{2\sigma_0^2}\right)}{\sigma_0 \sqrt{2\pi}} \prod_{i=1}^{n} \frac{\exp\left(-\frac{(Y_i - \mu)^2}{2\sigma^2}\right)}{\sigma \sqrt{2\pi}} \propto \exp\left(-\frac{(\mu - \hat{\mu}_0)^2}{2\hat{\sigma}_0^2}\right), \tag{17}$$

which is a Normal distribution, $N(\hat{\mu}_0, \hat{\sigma}_0)$, i.e. the same as the prior distribution with updated parameters

$$\mu_0 \to \hat{\mu}_0 = (\mu_0 + \omega \sum_{i=1}^{n} Y_i)/(1 + n \times \omega),$$
$$\sigma_0^2 \to \hat{\sigma}_0^2 = \sigma_0^2/(1 + n \times \omega), \quad \text{where } \omega = \sigma_0^2/\sigma^2. \tag{18}$$

The expected value of $Y_{n+1}$ (given past observations), $E[Y_{n+1} \mid \mathbf{X}]$, allows for a good interpretation, as follows:

$$E[Y_{n+1} \mid \mathbf{X}] = E[\mu \mid \mathbf{X}] = \hat{\mu}_0 = \frac{\mu_0 + \omega \sum_{i=1}^{n} Y_i}{1 + n \times \omega} = w\overline{Y} + (1 - w)\mu_0, \tag{19}$$

where

$\overline{Y} = \frac{1}{n} \sum_{i=1}^{n} Y_i$ is the estimate of $\mu$ using the observed losses only,

$\mu_0$ is the estimate of $\mu$ using a prior distribution only (e.g. specified by expert),

$w = \dfrac{n}{n + \sigma^2/\sigma_0^2}$ is the credibility weight in [0,1) used to combine $\mu_0$ and $\overline{Y}$.

As the number of observations increases, the credibility weight $w$ increases and vice versa. That is, the more observations we have the greater weight we assign to the estimator based on the observed counts and the lesser weight is attached to the expert opinion estimate. Also, larger uncertainty in the expert opinion $\sigma_0^2$ leads to a higher credibility weight for observations and larger volatility of observations $\sigma^2$ leads to a higher credibility weight for expert opinions.

The posterior distribution can be calculated recursively as follows. Consider observations $Y_1, Y_2, ..., Y_k, ...$. Assume that the prior distribution $\pi(\mu | \mu_0, \sigma_0)$, $N(\mu_0, \sigma_0)$, is specified initially, then the posterior distribution $\hat{\pi}_k(\mu | Y_1, ..., Y_k)$ after the $k$-th year is the Normal distribution $N((\hat{\mu}_0)_k, (\hat{\sigma}_0)_k)$ with

$$(\hat{\mu}_0)_k = (\mu_0 + \omega \sum_{i=1}^{k} Y_i)/(1 + k \times \omega) \text{ and } (\hat{\sigma}_0^2)_k = \sigma_0^2/(1 + k \times \omega),$$

where $\omega = \sigma_0^2/\sigma^2$. It is easy to show that

$$(\hat{\mu}_0)_k = \dfrac{(\hat{\mu}_0)_{k-1} + [(\hat{\sigma}_0^2)_{k-1}/\sigma^2] \times Y_k}{1 + [(\hat{\sigma}_0^2)_{k-1}/\sigma^2]}, \quad (\hat{\sigma}_0^2)_k = \dfrac{(\hat{\sigma}_0^2)_{k-1}}{1 + [(\hat{\sigma}_0^2)_{k-1}/\sigma^2]}. \tag{20}$$

That is, calculation of posterior distribution parameters can be based on the most recent observation and the parameters of the posterior distribution calculated just before this observation.

### 3.3 Pareto-Gamma (severity modelling)

Another important example of the severity distribution, which is very useful to fit the tail of the distribution, for a given threshold $L > 0$, is the Pareto distribution with a density

$$f(x | \xi) = \dfrac{\xi}{L} \left( \dfrac{x}{L} \right)^{-\xi - 1}. \tag{21}$$

It is defined for $x \geq L$ and $\xi > 0$. If $\xi > 1$, then the mean is $L\xi/(\xi - 1)$, otherwise the mean does not exist. Suppose that conditionally, given $\xi$, observations $\mathbf{X} = (X_1, ..., X_n)$ are independent random variables from the above Pareto distribution and the tail parameter $\xi$ has a prior Gamma distribution, $Gamma(\alpha, \beta)$, with a density

$$\pi(\xi \mid \alpha, \beta) \propto \xi^{\alpha-1} \exp(-\xi/\beta). \tag{22}$$

Using formula (5), the posterior distribution

$$\hat{\pi}(\xi \mid \mathbf{X}) = \xi^n \exp\left[-(\xi+1)\sum_{i=1}^{n}\ln\left(\frac{X_i}{L}\right)\right] \times \xi^{\alpha-1} \exp\left(-\frac{\xi}{\beta}\right) \propto \xi^{\hat{\alpha}-1} \exp\left[-\frac{\xi}{\hat{\beta}}\right] \tag{23}$$

is Gamma, $Gamma(\hat{\alpha}, \hat{\beta})$, i.e. the same as the prior distribution with updated parameters

$$\alpha \to \hat{\alpha} = \alpha + n,$$
$$\beta^{-1} \to \hat{\beta}^{-1} = \beta^{-1} + \sum_{i=1}^{n}\ln\left(\frac{X_i}{L}\right). \tag{24}$$

The mean of the posterior distribution for $\xi$ allows for a good interpretation, as follows:

$$\hat{\xi} = E[\xi \mid \mathbf{X}] = \hat{\alpha} \times \hat{\beta} = \frac{\alpha + n}{\frac{1}{\beta} + \sum_{i=1}^{n}\ln\left(\frac{X_i}{L}\right)} = w\hat{\xi}^{MLE} + (1-w)\xi_0, \tag{25}$$

where

$\hat{\xi}^{MLE} = \frac{1}{n}\sum_{i=1}^{n}\ln\left(\frac{X_i}{L}\right)$ is the maximum likelihood estimate of $\xi$ using the observed losses,

$\xi_0 = \alpha \times \beta$ is the estimate of $\xi$ using a prior distribution only (e.g. specified by expert),

$w = \left[\sum_{i=1}^{n}\ln\left(\frac{X_i}{L}\right)\right] \times \left[\sum_{i=1}^{n}\ln\left(\frac{X_i}{L}\right) + \frac{1}{\beta}\right]^{-1}$ is the weight in [0,1) combining $\xi_0$ and $\hat{\xi}^{MLE}$.

The posterior distribution can be easily calculated recursively. Consider observations $X_1, X_2, \ldots, X_k, \ldots$. Assume that the prior distribution $\pi(\xi \mid \alpha, \beta)$, $Gamma(\alpha, \beta)$, is specified initially, then the posterior distribution $\hat{\pi}_k(\xi \mid X_1, \ldots, X_k)$ after the $k$-th year is the Gamma distribution $Gamma(\hat{\alpha}_k, \hat{\beta}_k)$ with $\hat{\alpha}_k = \alpha + k$ and $\hat{\beta}_k^{-1} = \beta^{-1} + \sum_{i=1}^{k}\ln(X_i/L)$.

It is easy to show that

$$\hat{\alpha}_k = \hat{\alpha}_{k-1} + 1, \ \hat{\beta}_k^{-1} = \hat{\beta}_{k-1}^{-1} + \ln\left(\frac{X_k}{L}\right). \tag{26}$$

Again, this leads to a very efficient recursive scheme, where the calculation of the posterior distribution parameters is based on the most recent observation and parameters of the posterior distribution calculated just before this observation.

It is important to note that the prior and posterior distributions of $\xi$ are Gamma distributions formally defined for $\xi > 0$. Thus, there is a finite probability that $\Pr[\xi \leq 1] > 0$, which leads to infinite means of predicted distributions, i.e. $E[X_i] = \infty$ and $E[X_{n+1} | \mathbf{X}] = \infty$. If we do not want to allow for infinite mean behavior, then $\xi$ should be restricted to $\xi > 1$. In the following section we explain how to deal with such problems.

## *3.4 Restricted Parameters*

In practice, it is not unusual to restrict parameters. For example, for given observations $\mathbf{X} = (X_1, ..., X_n)$ we choose the LogNormal distribution, $LN(\mu, \sigma)$, and we choose a prior distribution for $\mu$ to be the Normal distribution, $N(\mu_0, \sigma_0)$. However, if we know that $\mu$ cannot be negative, we restrict $N(\mu_0, \sigma_0)$ to nonnegative values only. Another example is the Pareto-Gamma case, where the prior distribution for parameter $\xi$ is $Gamma(\alpha, \beta)$, defined for $\xi > 0$. But if we do not want to allow for infinite mean predicted loss, then the parameter should be restricted to $\xi > 1$. These cases can be easily handled by using the truncated versions of the prior-posterior distributions. Assume that $\pi(\theta)$ is a prior distribution whose posterior distribution is $\hat{\pi}(\theta | \mathbf{X}) = h(\mathbf{X} | \theta) \pi(\theta) / h(\mathbf{X})$, where $\theta$ is unrestricted. If the parameter is restricted to $a \leq \theta \leq b$, then we can consider the prior distribution

$$\pi^{tr}(\theta) = \frac{\pi(\theta)}{\Pr[a \leq \theta \leq b]} \times I_{a \leq \theta \leq b}, \quad \Pr[a \leq \theta \leq b] = \int_a^b \pi(\theta) d\theta, \qquad (27)$$

for some $a$ and $b$, with $\Pr[a \leq \theta \leq b] > 0$. Here: $I_{a \leq \theta \leq b} = 1$, if $a \leq \theta \leq b$, and equals zero otherwise. $\Pr[a \leq \theta \leq b]$ plays the role of normalization and thus the posterior distribution for that prior is simply

$$\hat{\pi}^{tr}(\theta | \mathbf{X}) = \frac{\hat{\pi}(\theta | \mathbf{X})}{\Pr[a \leq \theta \leq b | \mathbf{X}]} I_{a \leq \theta \leq b}, \quad \Pr[a \leq \theta \leq b | \mathbf{X}] = \int_a^b \hat{\pi}(\theta | \mathbf{X}) d\theta. \qquad (28)$$

It is obvious that if $\pi(\theta)$ is a conjugate prior then $\pi^{tr}(\theta)$ is a conjugate prior too.

# 4 Estimating structural prior parameters subjectively

In general, the structural parameters of the prior distributions can be estimated subjectively using expert opinions (pure Bayesian approach) and using data (empirical Bayesian approach). The latter will be considered in the next section. In a pure Bayesian approach, the prior distribution is specified subjectively (that is, in the context of operational risk, using expert opinions). Berger (1985) lists several methods (below, $\Theta$ is a parameter space of $\boldsymbol{\theta}$):

- **Histogram approach**: split $\Theta$ into intervals and specify the subjective probability for each interval. From this, the smooth density of the prior distribution can be determined.
- **Relative Likelihood Approach**: compare the intuitive likelihoods of the different points in $\Theta$. Again, the smooth density of prior distribution can be determined. It is difficult to apply this method in the case of unbounded parameters.
- **CDF determinations**: subjectively construct the cumulative distribution function for the prior and sketch a smooth curve.
- **Matching a Given Functional Form**: find the prior distribution parameters assuming some functional form for the prior distribution to match prior beliefs (on the moments, quantiles, etc) as close as possible.

Below, using the method of matching a given function form, we consider the estimation of the prior distribution parameters for Poisson-Gamma, Pareto-Gamma and LogNormal-Normal distribution pairs. The use of a particular method is determined by a specific problem and expert experience. Usually, if the expected values for the quantiles (or mean) and their uncertainties are estimated by the expert then it is possible to fit the priors.

## *4.1 Poisson-Gamma*

Suppose that the annual frequency of the operational risk losses $N$ is modeled by the Poisson distribution, $Poisson(\lambda)$, and the prior distribution $\pi(\lambda | \alpha, \beta)$ for $\lambda$ is $Gamma(\alpha, \beta)$. Then, $E[N | \lambda] = \lambda$ and $E[\lambda] = \alpha \times \beta$, see Section 3.1. The expert may estimate the expected number of events but he can not be certain in the estimate. One could say that the expert's "best" estimate for the expected number of events corresponds to $E[E[N | \lambda]] = E[\lambda]$. If the expert specifies $E[\lambda]$ and an uncertainty that the "true" $\lambda$ for next year is within the interval [$a$,$b$] with the probability $\Pr[a \le \lambda \le b] = p$ (it may be convenient to set $p = 2/3$), then the equations

$$E[\lambda] = \alpha \times \beta,$$
$$\Pr[a \le \lambda \le b] = p = \int_a^b \pi(\lambda | \alpha, \beta) d\lambda = F_{\alpha,\beta}^{(G)}[b] - F_{\alpha,\beta}^{(G)}[a] \tag{29}$$

can be solved numerically to estimate the structural parameters $\alpha$ and $\beta$. Here, $F_{\alpha,\beta}^{(G)}[.]$ is the cumulative Gamma distribution with parameters $\alpha$ and $\beta$

$$F_{\alpha,\beta}^{(G)}[y] = \int_0^y \frac{x^{\alpha-1}}{\Gamma(\alpha)\beta^\alpha} \exp\left(-\frac{x}{\beta}\right) dx. \tag{30}$$

In the insurance industry, the uncertainty for the "true" $\lambda$ is often measured in terms of the coefficient of variation, $Vco(\lambda) = \sqrt{Var(\lambda)}/E[\lambda]$. Given the expert estimates for $E[\lambda] = \alpha \times \beta$ and $Vco(\lambda) = 1/\sqrt{\alpha}$, the structural parameters $\alpha$ and $\beta$ are easily estimated.

For example: if the expert specifies $E[\lambda] = 0.5$ and $\Pr[0.25 \leq \lambda \leq 0.75] = 2/3$, then we can fit a prior distribution, $Gamma(\alpha \approx 3.407, \beta \approx 0.147)$ by solving (29). Assume now that the bank experienced no losses over the first year (after the prior distribution was estimated). Then, using formulas (13), the posterior distribution parameters are $\hat{\alpha}_1 \approx 3.407 + 0 = 3.407$, $\hat{\beta}_1 \approx 0.147/(1+0.147) \approx 0.128$ and the estimated arrival rate using the posterior distribution is $\hat{\lambda}_1 = \hat{\alpha}_1 \times \hat{\beta}_1 \approx 0.436$. If during the next year no losses are observed again, then the posterior distribution parameters are $\hat{\alpha}_2 = \hat{\alpha}_1 + 0 \approx 3.407$, $\hat{\beta}_2 = \hat{\beta}_1/(1+\hat{\beta}_1) \approx 0.113$ and $\hat{\lambda}_2 = \hat{\alpha}_2 \times \hat{\beta}_2 \approx 0.385$. Subsequent observations will update the arrival rate estimator correspondingly using formulas (13). Thus, starting from the expert specified prior, observations regularly update (refine) the posterior distribution. The expert might reassess a posterior distribution at any point in time (the posterior distribution can be treated as a prior distribution for the next period), if new practices/policies were introduced in the bank that affect the frequency of the loss. That is, if we have a new policy at time $k$, expert may reassess parameters and replace $\hat{\alpha}_k$ and $\hat{\beta}_k$ by $\hat{\alpha}_k^*$ and $\hat{\beta}_k^*$ respectively.

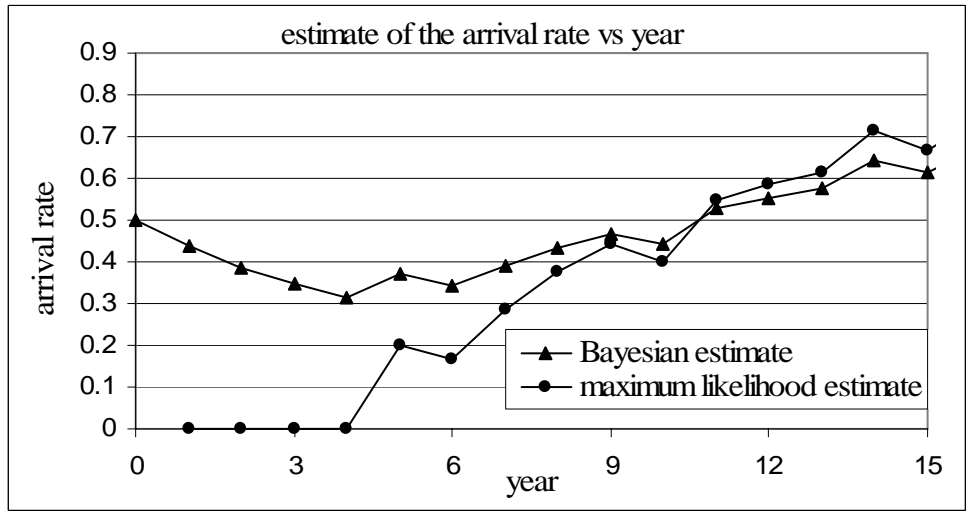

**Figure 1.** The Bayesian and the standard maximum likelihood estimates of the arrival rate vs the observation year. The Bayesian estimate is a mean of the posterior distribution when the prior distribution is Gamma with $\alpha \approx 3.41$ and $\beta \approx 0.15$. The maximum likelihood estimate is a simple average over the number of observed events. The annual counts (0, 0, 0, 0, 1, 0, 1, 1, 1, 0, 2, 1, 1, 2, 0) were sampled from the Poisson with $\lambda = 0.6$.

In Figure 1, we show the posterior best estimate for the arrival rate $\hat{\lambda}_k = \hat{\alpha}_k \times \hat{\beta}_k$, $k = 1,...,15$ (with the prior distribution as in the above example), when the annual number of events $N_k$, $k = 1,...,15$, are simulated from the Poisson distribution with $\lambda = 0.6$. On the same figure, we show the standard maximum likelihood estimate of the arrival rate $\tilde{\lambda}_k = \frac{1}{k}\sum_{i=1}^{k} N_i$. After approximately 8 years, the estimators are very close to each other. However, for a small number of observed years, the Bayesian estimate is more accurate as it takes the prior information into account. Only after 12 years, both estimators converge to the true value of 0.6 (this is because the bank was very lucky to have no events during the first four years). Note that for this example we assumed the prior distribution with a mean equal to 0.5, which is different from the true arrival rate. Thus, this example shows that an initially incorrect prior estimator is corrected by the observations as they become available. It is interesting to observe that, in year 14, the estimators become slightly different again. This is because the bank was unlucky to experience event counts 1, 1, and 2 in the years 12, 13, and 14 respectively. As a result, the maximum likelihood estimate becomes higher than the true value, while the Bayesian estimate is more stable (smooth) in respect to the unlucky years. If this example is repeated with different sequences of random numbers, then one would observe quite different maximum likelihood estimates (for small $k$) and more stable Bayesian estimates.

## *4.2 LogNormal-Normal*

Suppose that *X*, the severity of operational losses, is modeled by the LogNormal distribution, $LN(\mu,\sigma)$, see Section 3.2. Then, for given $\mu$ and $\sigma$, the expected loss is

$$E[X \mid \mu,\sigma] = M(\mu,\sigma) = \exp(\mu + \tfrac{1}{2}\sigma^2) \tag{31}$$

and the quantile at level *q* is

$$Q_q(\mu,\sigma) = \exp(\mu + \sigma Z_q), \tag{32}$$

where $Z_q$ is the standard Normal quantile at the level *q*. Consider the case when $\sigma$ is known and the prior distribution for $\mu$ is $N(\mu_0,\sigma_0)$. In this case, unconditionally, $M(\mu,\sigma)$ is distributed from $LN(\mu_0 + \tfrac{1}{2}\sigma^2,\sigma_0)$ and the quantile $Q_q(\mu,\sigma)$ is distributed from $LN(\mu_0 + \sigma Z_q,\sigma_0)$.

The expert may specify "the best" estimate of the expected loss $E[M(\mu,\sigma)]$ and uncertainty, i.e. the interval [*a,b*] such that the true expected loss is within the interval with a probability $p = \Pr[a \leq M \leq b]$. Then the equations

$$E[M] = \exp(\mu_0 + \tfrac{1}{2}\sigma^2 + \tfrac{1}{2}\sigma_0^2),$$
$$p = \Pr[a \leq M \leq b] = \Phi\left[\frac{\ln b - \tfrac{1}{2}\sigma^2 - \mu_0}{\sigma_0}\right] - \Phi\left[\frac{\ln a - \tfrac{1}{2}\sigma^2 - \mu_0}{\sigma_0}\right] \tag{33}$$

can be solved to find $\mu_0, \sigma_0$. Here, $\Phi[.]$ is the cumulative standard Normal distribution. For example, if $\sigma = 2$ and the expert estimates are $E[M] = 10$ and $p = \Pr[8 \leq M \leq 12] = 2/3$, then we can fit $\mu_0 \approx 0.28$ and $\sigma_0 \approx 0.21$ and use (20) to calculate the posterior parameters $(\hat\mu_0)_k, (\hat\sigma_0)_k$ as observations $X_k, k = 1,2,...$ become available.

Also, one can fit parameters $\mu_0$ and $\sigma_0$ using estimates for some quantile and uncertainty by solving

$$E[Q_q] = \exp(\mu_0 + \sigma Z_q + \tfrac{1}{2}\sigma_0^2),$$
$$p = \Pr[a \leq Q_q \leq b] = \Phi\left[\frac{\ln b - \sigma Z_q - \mu_0}{\sigma_0}\right] - \Phi\left[\frac{\ln a - \sigma Z_q - \mu_0}{\sigma_0}\right]. \tag{34}$$

If the uncertainty for $M(\mu,\sigma)$ or $Q_q(\mu,\sigma)$ in (33)-(34) is measured using the coefficient of variation $Vco(x) = \sqrt{Var(x)}/E[x]$, then $\mu_0, \sigma_0$ are easily expressed in the closed form. In the insurance industry $Vco$ is often provided by regulators.

If the prior distributions for both $\mu$ and $\sigma$ are required, see Appendix A, we can use a simple relationship between $\sigma$ and two quantiles $Q_{q(2)}, Q_{q(1)}$:

$$\sigma = \ln(Q_{q(2)}/Q_{q(1)})/(Z_{q(2)} - Z_{q(1)}). \tag{35}$$

Then, one can try to fit the prior distribution for $\sigma$ using the expert opinions on, e.g. $E[\ln(Q_{q(2)}/Q_{q(1)})]$ and $\Pr[a \leq Q_{q(2)}/Q_{q(1)} \leq b]$ or the opinions involving several pairs of quantiles. Given $\sigma$, the prior distribution for $\mu$ can be estimated using equations (33) or (34).

### *4.3 Pareto-Gamma*

Suppose that *X*, the severity of operational losses exceeding threshold *L*, is modeled by the Pareto distribution, *Pareto*$(\xi)$. Then, for given $\xi$, $E[X|\xi] = \mu(\xi) = L\xi/(\xi-1)$, with $\xi > 1$, and the quantile at level $q$ is $Q_q(\xi) = L\exp[-\ln(1-q)/\xi]$, with $\xi > 0$, see Section 3.3. It is reasonable to assume that, unconditionally, expected loss is finite, then the tail parameter $\xi$ should satisfy $\xi \geq B > 1$ and we can choose the prior distribution for $\xi$ to be a Gamma distribution truncated below *B*

$$\pi(\xi|\alpha,\beta) = I_{B \leq \xi} \times \frac{\xi^{\alpha-1}\exp(-\xi/\beta)}{(1 - F_{\alpha,\beta}^{(G)}[B]) \times \Gamma(\alpha)\beta^\alpha}, \quad \xi \geq B, \alpha > 0, \beta > 0, \tag{36}$$

where $F_{\alpha,\beta}^{(G)}[.]$ is a cumulative Gamma distribution. If the expert estimates $E[\xi]$ and the uncertainty $\Pr[a \leq \xi \leq b] = p$, then the following two equations

$$E[\xi] = \alpha \times \beta \times \frac{1 - F^{(G)}_{\alpha+1,\beta}[B]}{1 - F^{(G)}_{\alpha,\beta}[B]},$$

$$\Pr[a \leq \xi \leq b] = \frac{F^{(G)}_{\alpha,\beta}[b] - F^{(G)}_{\alpha,\beta}[a]}{1 - F^{(G)}_{\alpha,\beta}[B]} \tag{37}$$

can be solved to estimate the structural parameters $\alpha$ and $\beta$. Assume that, the lower bound for the tail parameter is $B=2$ and the expert estimates are $E[\xi]=5$, $\Pr[4 \leq \xi \leq 6] = 2/3$. Then we can fit $\alpha \approx 23.086$, $\beta \approx 0.217$ and can calculate the posterior distribution parameters $\hat{\alpha}_k, \hat{\beta}_k$, when observations $X_k, k=1,2,...$, become available, using (26). In Figure 2, we show the subsequent posterior best estimates for the tail parameter

$$\hat{\xi}_k = \hat{\alpha}_k \times \hat{\beta}_k \times (1 - F^{(G)}_{\hat{\alpha}+1,\hat{\beta}}[B])/(1 - F^{(G)}_{\hat{\alpha},\hat{\beta}}[B]), \; k=1,2,..., \tag{38}$$

when the losses $X_k$ are simulated from the Pareto distribution with $\xi=4$ and $L=1$. On the same figure, we show the standard maximum likelihood estimate of the tail parameter $\tilde{\xi}_k = [\frac{1}{k}\sum_{i=1}^{k} \ln(X_i/L)]^{-1}$. It is easy to see that the Bayesian estimates are more stable while the maximum likelihood estimates are quite volatile when the number of observations is small. As the number of observations increases, two estimators become almost the same.

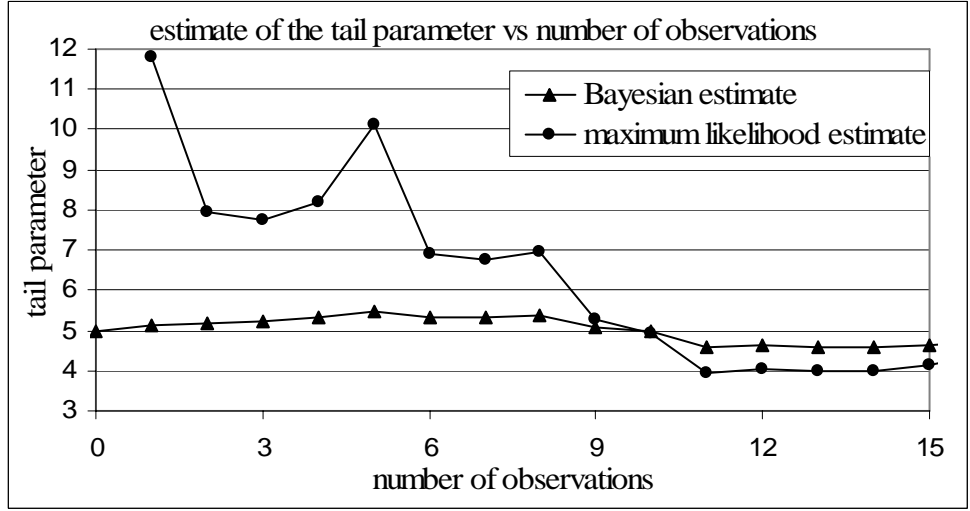

**Figure 2.** The Bayesian and the standard maximum likelihood estimates of the Pareto tail parameter vs the number of observations. The Bayesian estimate is a mean of posterior distribution when the prior distribution is Gamma with $\alpha \approx 23.1$, $\beta \approx 0.22$, truncated below $B=2$. The actual losses were sampled from the Pareto distribution with $\xi=4$ and $L=4$.

If it is difficult to express opinions on $\xi$ directly, then the expert may try to estimate expected loss, quantile or their uncertainties. It might be difficult numerically to fit $\alpha,\beta$ if expert specifies expected loss $E[\mu(\xi)]$ or expected quantile $E[Q_q(\xi)]$

$$E[\mu(\xi)] = \int_B^\infty \mu(\xi)\pi(\xi\,|\,\alpha,\beta)d\xi, \quad E[Q_q(\xi)] = L\int_B^\infty Q_q(\xi)\pi(\xi\,|\,\alpha,\beta)d\xi, \qquad (39)$$

as these are not easily expressed, although there is no problem in principle. Fitting opinions on uncertainties might be easier. For example, if the expert estimates the interval $[a,b]$ such that the true expected loss is within the interval with the probability $\Pr[a \le \mu(\xi) \le b] = p$ then it leads to the equation

$$\Pr[a \le \mu(\xi) \le b] = p = \int_{\tilde{b}}^{\tilde{a}} \pi(\xi\,|\,\alpha,\beta)d\xi = \frac{F^{(G)}_{\alpha,\beta}[\tilde{a}] - F^{(G)}_{\alpha,\beta}[\tilde{b}]}{1 - F^{(G)}_{\alpha,\beta}[B]}, \quad \tilde{a} = \frac{a}{a-L}, \tilde{b} = \frac{b}{b-L}. \qquad (40)$$

Here, the interval bounds should satisfy $L < a < b \le B \times L/(B-1)$. The estimation of the interval $[a,b], L < a < b$, such that the true quantile is within the interval with the probability $\Pr[a \le Q_q(\xi) \le b] = p$ leads to the equation

$$\Pr[a \le Q_q(\xi) \le b] = L\int_{C_1}^{C_2} \pi(\xi\,|\,\alpha,\beta)d\xi = \frac{F^{(G)}_{\alpha,\beta}[C_2] - F^{(G)}_{\alpha,\beta}[C_1]}{1 - F^{(G)}_{\alpha,\beta}[B]},$$
$$C_1 = -\frac{\ln(1-q)}{\ln(b/L)}, \quad C_2 = -\frac{\ln(1-q)}{\ln(a/L)}, \qquad (41)$$

where the interval bounds should satisfy $L < a < b \le L\exp(-[\ln(1-q)]/B)$. Equations (40) and (41) or similar ones can be used to fit $\alpha$ and $\beta$. If the expert specifies more than two quantities, then one can use, for example, a nonlinear least square procedure to fit the structural parameters.

## 5 Estimation of the prior parameters using data

The prior distribution can be estimated using a marginal distribution of observations. The data can be collective industry data, collective data in the bank, etc.

**The maximum likelihood estimator**.
For example, consider a specific risk cell (event type/business line) in $J$ banks with the observations $X_{j,k}$, $k = 1,...,K_j$, $j = 1,...,J$. Here, $K_j$ is the number of observations in bank $j$. Assume that, $X_{j,k}, k = 1,...,K_j$ are conditionally independent and identically distributed from $f(X_{j,k}\,|\,\boldsymbol{\theta}_j)$, for given $\boldsymbol{\theta}_j$. That is, the risk cell in the $j$-th bank has its own risk profile $\boldsymbol{\theta}_j$. Assume now that $\boldsymbol{\theta}_j$, $j = 1,...,J$, are independent and identically distributed from $\pi(\boldsymbol{\theta}_j)$, that is we assumed that the risk cells in different banks are the same a priori (before we have any observations). Then the likelihood of all observations can be written as

$$\Psi = \prod_{j=1}^{J} \int \left[ \prod_{k=1}^{K_j} f(X_{j,k} | \boldsymbol{\theta}_j) \right] \pi(\boldsymbol{\theta}_j) d\boldsymbol{\theta}_j . \tag{42}$$

The parameters of $\pi(\boldsymbol{\theta}_j)$ can be estimated by maximizing the above likelihood. The distribution $\pi(\boldsymbol{\theta}_j)$ is a prior distribution for the cell in the *j*-th bank. Using internal data of the risk cell in the *j*-th bank, its posterior distribution is calculated using (5) as

$$\hat{\pi}(\boldsymbol{\theta}_j | X_{j,k}, k = 1,...,K_j) = \prod_{k=1}^{K_j} f(X_{j,k} | \boldsymbol{\theta}_j) \pi(\boldsymbol{\theta}_j) , \tag{43}$$

where $\pi(\boldsymbol{\theta}_j)$ was fitted with MLE in (42). The basic idea here is that the maximum likelihood estimates based on observations from all banks are better then those obtained using smaller number of observations available in the risk cell of a particular bank.

**The maximum likelihood estimator with a priori differences.**
It is not difficult to include a priori known differences (for example, exposure indicators, expert opinions on the differences, etc) between the risk cells from the different banks. As an example, we consider the case when the annual frequency of the events is modeled by the Poisson distribution with the Gamma prior and estimate structural parameters using the industry data with differences between the banks taken into account. Consider a risk cell in *J* banks with observations $N_{j,k}$, $k = 1,...,K_j$, $j = 1,...,J$. So, $N_{j,k}$ is the annual number of events observed in the cell of the *j*-th bank in the *k*-th year. Also, denote $\mathbf{N}_j = (N_{j1},...,N_{j,K_j})$. Assume that $N_{j,k}$ are conditionally independent and identically distributed from $f(N_{j,k} | \lambda_j) = Poisson(\lambda_j \times V_{j,k})$, for given $\lambda_j$. Here, $V_{j,k}$ is the known constant (i.e. the gross income or the volume or combination of several exposure indicators) and $\lambda_j$ is a risk profile of the cell in the *j*-th bank. Also, assume that $\lambda_j$, $j = 1,...,J$ are independent and identically distributed from $\pi(\lambda_j) = Gamma(\alpha, \beta)$. Denote $N_j = \sum_k^{K_j} N_{j,k}$, $V_j = \sum_k^{K_j} V_{j,k}$. Then, similar to (42), the likelihood of observations can be written as

$$\Psi = \prod_{j=1}^{J} \int \left[ \prod_{k=1}^{K_j} f(N_{j,k} | \lambda_j) \right] \pi(\lambda_j) d\lambda_j$$

$$= \prod_{j=1}^{J} \int \left[ \prod_{k=1}^{K_j} e^{-\lambda_j V_{j,k}} \frac{(V_{j,k} \lambda_j)^{N_{j,k}}}{(N_{j,k})!} \right] \times \frac{\lambda_j^{\alpha-1} e^{-\lambda_j/\beta}}{\Gamma(\alpha) \beta^{\alpha}} d\lambda_j \qquad (44)$$

$$= \left[ \prod_{j=1}^{J} \prod_{k=1}^{K_j} \frac{(V_{j,k})^{N_{j,k}}}{(N_{j,k})!} \right] \times \prod_{j=1}^{J} \frac{\Gamma(\alpha + N_j)}{\Gamma(\alpha) \beta^{\alpha} (V_j + 1/\beta)^{\alpha+N_j}}.$$

The parameters can now be estimated by maximizing the log-likelihood

$$\ln \Psi \propto \sum_{j=1}^{J} \left\{ \ln \Gamma(\alpha + N_j) - \ln \Gamma(\alpha) - \alpha \ln \beta - (\alpha + N_j) \ln\left(\frac{1}{\beta} + V_j\right) \right\}. \qquad (45)$$

To avoid the use of numerical optimization required for maximizing (45), one could also use a method of moments. Denote $\lambda_0 = E[\lambda_j] = \alpha \beta$, $\sigma_0^2 = \text{Var}[\lambda_j] = \alpha \beta^2$. The appropriate estimators $\hat{\lambda}_0$ and $\hat{\sigma}_0^2$ for $\lambda_0$ and $\sigma_0^2$ respectively are

$$\hat{\lambda}_0 = \frac{1}{J} \sum_{j=1}^{J} \hat{\lambda}_j, \quad \hat{\sigma}_0^2 = \max\left[ \frac{1}{J-1} \sum_{j=1}^{J} (\hat{\lambda}_j - \hat{\lambda}_0)^2 - \frac{\hat{\lambda}_0}{J} \sum_{j=1}^{J} \frac{1}{K_j^2} \sum_{k=1}^{K_j} \frac{1}{V_{j,k}}, 0 \right],$$

$$\text{where } \hat{\lambda}_j = \frac{1}{K_j} \sum_{k=1}^{K_j} \frac{N_{j,k}}{V_{j,k}}, \; j=1,...,J, \qquad (46)$$

see Appendix B for proof. These can easily be used to estimate $\alpha$ and $\beta$ as $\hat{\alpha} = \hat{\lambda}_0 / \hat{\beta}$ and $\hat{\beta} = \hat{\sigma}_0^2 / \hat{\lambda}_0$ correspondingly.

Once the prior distribution parameters $\alpha$ and $\beta$ are estimated, then, using (5), the posterior distribution of $\lambda_j$, for the $j$-th bank is

$$\hat{\pi}(\lambda_j | \mathbf{N}_j) \propto \frac{(\lambda_j/\beta)^{\alpha-1}}{\Gamma(\alpha)\beta} e^{-\lambda_j/\beta} \prod_{k=1}^{K_j} e^{-\lambda_j V_{j,k}} \frac{(V_{j,k}\lambda_j)^{N_{j,k}}}{N_{j,k}!} \propto \lambda^{N_j + \alpha - 1} \exp\left(-\lambda_j V_j - \frac{\lambda_j}{\beta}\right), \quad (47)$$

which is $Gamma(\hat{\alpha}, \hat{\beta})$ with

$$\hat{\alpha} = \alpha + \sum_{k=1}^{K_j} N_{j,k} \text{ and } \hat{\beta} = \beta / (1 + \beta \times \sum_{k=1}^{K_j} V_{j,k}). \qquad (48)$$

Assume that, the exposure indicator of the cell in the $j$-th bank for the next year is $V_{j,K_j+1} = V$. Then, the predictive distribution for the annual number of events in the cell (conditional on the past internal data) is Negative Binomial, $NegBin(\hat{\alpha}, \hat{p} = 1/(1+V\hat{\beta}))$:

$$f(N_{K_j+1} = N | \mathbf{N}_j) = \int e^{-\lambda V} \frac{(V\lambda)^N}{N!} \frac{\lambda^{\hat{\alpha}-1}}{\Gamma(\hat{\alpha})\hat{\beta}^{\hat{\alpha}}} e^{-\lambda/\hat{\beta}} d\lambda = \frac{\Gamma(N+\hat{\alpha})}{\Gamma(\hat{\alpha})N!}(1-\hat{p})^N \hat{p}^{\hat{\alpha}}. \quad (49)$$

Observe that we have scaled the parameters for considering a priori differences. This leads to a linear volume relation for the variance function, see (56) below. To obtain different functional relations, it might be better to scale the actual observations. For example, given observations $X_{j,k}$, $j = 1,...,J$, $k = 1,...,K_j$ (these could be frequencies or severities), consider variables $Y_{j,k} = X_{j,k}/V_{j,k}$. Assume that, for given $\boldsymbol{\theta}_j$, $Y_{j,k}$, $k = 1,...,K_j$ are independent and identically distributed from $f(.|\boldsymbol{\theta}_j)$. Also, assume that $\boldsymbol{\theta}_1,...,\boldsymbol{\theta}_J$ are independent and identically distributed from $\pi(.)$. Then one can construct the likelihood of $Y_{j,k}$ using (42) to fit parameters of $\pi(.)$ or try to use the method of moments.

# 6 The capital calculations and the discussion of dependence

For the purposes of the regulatory capital calculations of operational risk, the annual loss distribution (in particular its 0.999 quantile as a risk measure) should be quantified for each risk cell (event type/business line) in the bank. Consider a risk cell $j$ in the bank. Assume that the frequency $P_j(.|\boldsymbol{\lambda}_j)$ and severity $f_j(.|\boldsymbol{\alpha}_j)$ distributions, given $\boldsymbol{\lambda}_j$ and $\boldsymbol{\alpha}_j$, for the cell are chosen. Also, suppose that the posterior distributions $\hat{\pi}(\boldsymbol{\lambda}_j | \mathbf{N})$ and $\hat{\pi}(\boldsymbol{\alpha}_j | \mathbf{X})$ of $\boldsymbol{\lambda}_j$ and $\boldsymbol{\alpha}_j$ respectively are estimated using the prior distributions (for example, quantified via expert opinions or external data) weighted with the observed data using (5). Then, under the model (1), the annual loss distribution of the cell can be calculated using, for example, the Monte Carlo procedure with the following logical steps (where all random samples are independent):

**Step1**. For a given risk $j$, simulate the risk parameters $\boldsymbol{\lambda}_j$ and $\boldsymbol{\alpha}_j$ from their posterior distributions $\hat{\pi}(\boldsymbol{\lambda}_j | \mathbf{N})$ and $\hat{\pi}(\boldsymbol{\alpha}_j | \mathbf{X})$.

**Step2**. Given $\boldsymbol{\lambda}_j$ from the Step 1, simulate the annual number of events $N_j$ from the frequency distribution $P_j(.|\boldsymbol{\lambda}_j)$ of the $j$-th risk.

**Step3**. Given $\boldsymbol{\alpha}_j$ from the Step 1, simulate severities $X_{j,n}$, $n = 1,...,N_j$ from the severity distribution $f_j(.|\boldsymbol{\alpha}_j)$ of the $j$-th risk. Note that all severities $X_{j,n}$, $n = 1,...,N_j$ are simulated from the same distribution (i.e. risk profile parameter $\boldsymbol{\alpha}^{(j)}$ is applied for the whole year). This is because one of the assumptions in model (1) is that the severities are independent and identically distributed for given $\boldsymbol{\lambda}_j$ and $\boldsymbol{\alpha}_j$.

**Step 4**. Find the annual loss of the *j*-th risk as $Z_j = \sum_{n=1}^{N_j} X_{j,n}$.

**Step 5**. Repeat Steps 1-4 $K$ times to build a sample of the annual losses $Z_j(k)$, $k = 1,...,K$. Then, the 0.999 quantile (and other distribution characteristics if required) is estimated using the sample $Z_j(k), k = 1,...,K$ in a usual way.

If we assume that there are $J$ risk cells in the bank and these are independent, then the bank total annual loss can be calculated by repeating Steps 1-4 for each risk cell to find $Z_j$, $j = 1,...,J$. Then the bank annual loss is simply $Z_{tot} = \sum_{j}^{J} Z_j$. Repeating the whole procedure $K$ times will build a sample $Z_{tot}(k)$, $k = 1,...,K$, that can be used to estimate the annual loss distribution for the whole bank (and its capital). However, according to the Basel II requirements, see BIS (2005), the final bank capital should be calculated as a sum of the risk measures in the risk cells if bank's model cannot account for correlations between risks accurately. If this is the case, then Steps 1-5 should be performed for each risk cell to estimate their risk measures separately. Then these are summed to estimate the total bank capital. Of course, adding quantiles over the risk cells to find the quantile of the total loss distribution is too conservative as it is equivalent to the assumption of perfect dependence between risks.

An attractive way to model the dependence between risks is via dependence between risk profiles $\boldsymbol{\lambda}_j$, $\boldsymbol{\alpha}_j$, $j = 1,...,J$. This can be used to model the dependence between frequencies, between severities, and even between frequencies and severities. For example, the procedure can be as follows:

**Step 1**. Simulate the risk parameters $\boldsymbol{\lambda}_j$, $\boldsymbol{\alpha}_j$, $j = 1,...,J$ simultaneously from a multivariate distribution with the margins $\hat{\pi}(\boldsymbol{\lambda}_j | \mathbf{N})$, $\hat{\pi}(\boldsymbol{\alpha}_j | \mathbf{X})$ and an appropriate dependence structure (copula). For further information on the application of the copula method in finance, we refer to McNeil, Frey and Embrechts (2005).

**Step 2**. Given $\boldsymbol{\lambda}_j$ from the Step 1, simulate the annual number of events $N_j$ from the frequency distribution $P_j(.|\boldsymbol{\lambda}_j)$ for each risk cell $j = 1,...,J$.

**Step 3**. Given $\boldsymbol{\alpha}_j$ from the Step 1, and $N_j$ from Step 2, simulate independent severities $X_{j,n}, n = 1,...,N_j$ from the severity distribution $f_j(.|\boldsymbol{\alpha}_j)$ for each risk cell $j = 1,...,J$.

**Step 4**. Find the annual loss for each risk cell as $Z_j = \sum_{n=1}^{N_j} X_{j,n}$, $j = 1,...,J$, and the bank total loss as $Z_{tot} = \sum_{j}^{J} Z_j$.

**Step 5**. Repeat Steps 1-4 $K$ times to build a sample of the annual losses $Z_j(k)$, $k = 1,...,K$ and $Z_{tot}(k)$. Then, the 0.999 quantile (and other distribution characteristics if required) is estimated using the samples $Z_j(k), k = 1,...,K$ and $Z_{tot}(k)$ in a usual way.

If the risk profiles $\lambda_j$, $\alpha_j$, $j = 1,...,J$, are dependent one may discuss whether we determine the posterior distribution in the *j*-th risk cell using the data from the *j*-th risk cell only or using the data from all risk cells. That is, if there are two dependent risk cells *i* and *j* then we can learn something about posterior distribution in the *i*-th risk cell from the observations in the *j*-th risk cell and vice versa.

Note that in the above procedures we simulated the risk profiles $\lambda_j$, $\alpha_j$, $j = 1,...,J$, from their posterior distributions for each simulation. Thus, we model both the process uncertainty, which comes from the fact that $Z_j$ are random variables, and the parameter uncertainty, which comes from the fact that we do not know the true values of $\lambda_j$, $\alpha_j$.

The modelling of common events (shocks) that affect many risk cells simultaneously is an important part of operational risk modelling. These will introduce additional dependence between frequencies in the risk cells, see Lindskog and McNeil (2003).

Accurate quantification of the dependencies between the risks is a difficult task, which is an open field for future research.

# 7 Conclusions

In this paper we proposed to use the Bayesian inference method for the quantification of the frequency and severity distributions of operational risks. The method is based on specifying the prior distributions for the parameters of the frequency and severity distributions using expert opinions or industry data. Then, the prior distributions are weighted with the actual observations in the bank to estimate the posterior distributions of the model parameters. These are used to estimate the annual loss distribution for the next accounting year. The estimation of low frequency risks using this method has several appealing features such as: stable estimators, simple calculations (in the case of conjugate priors), and the ability to take into account expert opinions and industry data.

There are many other aspects of the Bayesian inference method that might be useful for operational risk modelling as well as related issues. For example, the hierarchical Bayesian approach can be used to estimate the prior distribution by combining several expert opinions with external data. A "toy" model is studied in Bühlmann, Shevchenko and Wüthrich (2006). The dependence between risks can potentially be introduced by considering, for example, evolutionary models, where the structural parameters (risk profiles) are evolving in time and are dependent.

One of the features of the described method is that the variance of the posterior distribution $\hat{\pi}(\theta|.)$ will converge to zero for a large number of observations. This means that the true value of the risk profile will be known exactly. However, there are many factors (for example, political, economical, legal, etc.) changing in time that should not allow for the precise knowledge of the risk profiles. One can model this by limiting the variance of the posterior distribution by some lower levels (e.g. 5%). This has been done in many solvency approaches for the insurance industry, see e.g. Swiss Solvency Test (2005), formulas (25)-(26).

In conclusion, we are very grateful to Paul Embrechts for his support and encouragement and Richard Jarrett for valuable comments.

## Appendix A
**The LogNormal distribution with the joint prior for both $\mu$ and $\sigma$.**

Suppose that conditionally, given $\mu$ and $\sigma$, observations $\mathbf{X} = (X_1,..., X_n)$ are independent and identically distributed random variables from the LogNormal distribution $LN(\mu,\sigma)$ with a density

$$f(x \mid \mu, \sigma) = \frac{1}{x\sqrt{2\pi\sigma^2}} \exp\left(-\frac{(\ln x - \mu)^2}{2\sigma^2}\right). \tag{50}$$

That is, $Y_i = \ln X_i$, $i = 1,...,n$ are independent and identically distributed from the Normal distribution $N(\mu,\sigma)$. Assume that the prior distribution of $\sigma^2$ is the Inverse Chi-squared distribution, $InvChiSq(\nu, \beta)$, and the prior distribution of $\mu$ (given $\sigma^2$) is the Normal distribution $N(\theta, \sigma^2/\phi)$ with the densities:

$$\pi(\sigma^2) = InvChiSq(\nu,\beta) = \frac{2^{-\nu/2}}{\beta\Gamma(\nu/2)}\left(\frac{\sigma^2}{\beta}\right)^{-\frac{\nu}{2}-1} \exp\left(-\frac{\beta}{2\sigma^2}\right),$$

$$\pi(\mu \mid \sigma^2) = N(\theta, \sigma^2/\phi) = \frac{1}{\sqrt{2\pi\sigma^2/\phi}} \exp\left(-\frac{(\mu-\theta)^2}{2\sigma^2/\phi}\right). \tag{51}$$

Then the joint prior distribution is

$$\pi(\mu, \sigma^2) = \frac{1}{\sqrt{2\pi\sigma^2/\phi}} \exp\left(-\frac{(\mu-\theta)^2}{2\sigma^2/\phi}\right) \times \frac{2^{-\nu/2}}{\beta\Gamma(\nu/2)}\left(\frac{\sigma^2}{\beta}\right)^{-\frac{\nu}{2}-1} \exp\left(-\frac{\beta}{2\sigma^2}\right)$$

$$\propto (\sigma^2)^{-\frac{\nu+1}{2}-1} \exp\left(-\frac{1}{2\sigma^2}[\beta + \phi(\mu-\theta)^2]\right). \tag{52}$$

It is easy to show that the marginal prior distribution for $\mu$ is shifted $t$-distribution with $\nu$ degrees of freedom as follows:

$$\pi(\mu) = \int \pi(\mu,\sigma^2)d\sigma^2 \propto \int x^{-\frac{\nu+1}{2}-1} \exp\left(-\frac{1}{2x}[\beta + \phi(\mu-\theta)^2]\right) dx$$

$$\propto \int y^{\frac{\nu+1}{2}-1} \exp\left(-\frac{y}{2}[\beta + \phi(\mu-\theta)^2]\right) dy \propto [\beta + \phi(\mu-\theta)^2]^{-\frac{\nu+1}{2}} \int z^{\frac{\nu+1}{2}-1} \exp(-z)dz \tag{53}$$

$$\propto [1 + \phi\nu(\mu-\theta)^2/(\nu\beta)]^{-\frac{\nu+1}{2}}.$$

Denote $\Psi(\sigma) = (\sigma^2)^{-\frac{\nu+1+n}{2}-1}$, $\bar{Y} = \frac{1}{n}\sum_{i=1}^{n} Y_i$ and $\overline{Y^2} = \frac{1}{n}\sum_{i=1}^{n} Y_i^2$. Then, the joint posterior distribution

$$\hat{\pi}(\mu, \sigma^2 \mid \mathbf{X}) \propto \Psi(\sigma)\exp\left(-\frac{1}{2\sigma^2}\left[\beta + \phi \times (\mu - \theta)^2 + \sum_{i=1}^{n}(Y_i - \mu)^2\right]\right)$$

$$\propto \Psi(\sigma)\exp\left(-\frac{1}{2\sigma^2}[\beta + (\phi + n)\mu^2 + \phi\theta^2 - 2\mu(\phi\theta + n\bar{Y}) + n\overline{Y^2}]\right)$$

$$\propto \Psi(\sigma)\exp\left(-\frac{1}{2\sigma^2}\left(\beta + \phi\theta^2 + n\overline{Y^2} - \frac{(\phi\theta + n\bar{Y})^2}{\phi + n} + (\phi + n)\left[\mu - \frac{\phi\theta + n\bar{Y}}{\phi + n}\right]^2\right)\right) \quad (54)$$

$$\propto (\sigma^2)^{-\frac{\hat{\nu}+1}{2}-1}\exp\left(-\frac{1}{2\sigma^2}[\hat{\beta} + \hat{\phi}(\mu - \hat{\theta})^2]\right)$$

has the same form as the joint prior distribution (52) with parameters updated as follows

$$\begin{aligned}
\nu &\to \hat{\nu} = \nu + n, \\
\beta &\to \hat{\beta} = \beta + \phi\theta^2 + n\overline{Y^2} - \frac{(\phi\theta + n\bar{Y})^2}{\phi + n}, \\
\theta &\to \hat{\theta} = \frac{\phi\theta + n\bar{Y}}{\phi + n}, \\
\phi &\to \hat{\phi} = \phi + n.
\end{aligned} \quad (55)$$

## Appendix B
**The method of moments to estimate the structural parameters in Section 5.**
Assume that, given $\lambda_j$, observations $N_{j,k}, k = 1,\ldots,K_j$, are independent and identically distributed from $Poisson(\lambda_j V_{j,k})$. Also, assume that $\lambda_j$, $j = 1,\ldots,J$, are independent and identically distributed from $Gamma(\alpha, \beta)$.

Denote $\lambda_0 = E[\lambda_j] = \alpha\beta$, $\sigma_0^2 = \text{Var}[\lambda_j] = \alpha\beta^2$ and consider the standardized frequencies $F_{j,k} = N_{j,k}/V_{j,k}$. It is easy to observe that,

$$\begin{aligned}
E[N_{j,k} \mid \lambda_j] &= \lambda_j V_{j,k}, \quad \text{Var}(N_{j,k} \mid \lambda_j) = \lambda_j V_{j,k}, \\
E[F_{j,k} \mid \lambda_j] &= \lambda_j, \quad \text{Var}(F_{j,k} \mid \lambda_j) = \lambda_j / V_{j,k}
\end{aligned} \quad (56)$$

and

$$E[F_{j,k}] = E[E[F_{j,k} | \lambda_j]] = E[\lambda_j] = \lambda_0,$$

$$\text{Var}(F_{j,k}) = E[\text{Var}(F_{j,k} | \lambda_j)] + \text{Var}(E[F_{j,k} | \lambda_j]) = E[\lambda_j / V_{j,k}] + \text{Var}[\lambda_j] = \frac{\lambda_0}{V_{j,k}} + \sigma_0^2. \quad (57)$$

Observe that $F_{j,k}$ are conditionally, given $\lambda_j$, independent and consider estimators $\hat{\lambda}_j = \frac{1}{K_j} \sum_{k=1}^{K_j} F_{j,k}, j = 1,...,J$. These estimators are independent and

$$E[\hat{\lambda}_j] = \frac{1}{K_j} \sum_{k=1}^{K_j} E[F_{j,k}] = \lambda_0,$$

$$\text{Var}(\hat{\lambda}_j) = E[\text{Var}(\hat{\lambda}_j | \lambda_j)] + \text{Var}(E[\hat{\lambda}_j | \lambda_j]) = \frac{\lambda_0}{K_j^2} \sum_{k=1}^{K_j} \frac{1}{V_{j,k}} + \sigma_0^2. \quad (58)$$

Thus

$$\hat{\lambda}_0 = \frac{1}{J} \sum_{j=1}^{J} \hat{\lambda}_j \quad (59)$$

is an unbiased estimator for $\lambda_0$. In the case of the same number of observations per company and the same weights $V_{j,k}$, this estimator would have a minimal variance among all linear combinations of $\hat{\lambda}_1,...,\hat{\lambda}_J$. Next, calculate

$$\sum_{j=1}^{J} E[(\hat{\lambda}_j - \hat{\lambda}_0)^2] = \sum_{j=1}^{J} E[(\hat{\lambda}_j - \lambda_0 + \lambda_0 - \hat{\lambda}_0)^2]$$

$$= \sum_{j=1}^{J} [\text{Var}(\hat{\lambda}_j) + \text{Var}(\hat{\lambda}_0) - 2\text{cov}(\hat{\lambda}_j, \hat{\lambda}_0)] = \sum_{j=1}^{J} [\text{Var}(\hat{\lambda}_j) + \frac{1}{J^2} \sum_{i=1}^{J} \text{Var}(\hat{\lambda}_i) - \frac{2}{J} \text{Var}(\hat{\lambda}_j)] \quad (60)$$

$$= \frac{J-1}{J} \sum_{j=1}^{J} \text{Var}(\hat{\lambda}_j) = \lambda_0 \frac{J-1}{J} \sum_{j=1}^{J} \frac{1}{K_j^2} \sum_{k=1}^{K_j} \frac{1}{V_{j,k}} + \sigma_0^2 (J-1).$$

Thus

$$\tilde{\sigma}_0^2 = \frac{1}{J-1} \sum_{j=1}^{J} (\hat{\lambda}_j - \hat{\lambda}_0)^2 - \frac{\hat{\lambda}_0}{J} \sum_{j=1}^{J} \frac{1}{K_j^2} \sum_{k=1}^{K_j} \frac{1}{V_{j,k}} \quad (61)$$

is an unbiased estimator for $\sigma_0^2$. Observe that $\tilde{\sigma}_0^2$ is not necessarily positive, hence we take $\hat{\sigma}_0^2 = \max\{\tilde{\sigma}_0^2, 0\}$ as the final estimator for $\sigma_0^2$.